\documentclass{jfm}
\usepackage{graphicx}
\usepackage{epstopdf, epsfig}
\usepackage{graphicx}
\usepackage{dcolumn}
\usepackage{bm}
\usepackage{siunitx}
\usepackage{xcolor}
\usepackage{amsmath,amssymb}
\usepackage{upgreek}

\graphicspath{{figs/}}

\shorttitle{Yielding to percolation: a universal scale}
\shortauthor{E. Chaparian}

\title{Yielding to percolation: a universal scale}

\author{Emad Chaparian
  \corresp{\email{emad.chaparian@strath.ac.uk}}}

\affiliation{James Weir Fluid Laboratory, Department of Mechanical \& Aerospace Engineering,\\ University of Strathclyde, Glasgow, United Kingdom}

\begin{document}

\maketitle

\begin{abstract}
A theoretical and computational study analysing the initiation of yield-stress fluids percolation in porous media is presented. Yield-stress fluid flows through porous media are complicated due to the non-linear rheological behaviour of this type of fluids, rendering the conventional Darcy type approach invalid. A critical pressure gradient must be exceeded to commence the flow of a yield-stress fluid in a porous medium. As the first step in generalising the Darcy law for yield-stress fluids, a universal scale based on the variational formulation of the energy equation is derived for the critical pressure gradient which reduces to purely geometrical feature of the porous media. The presented scaling is then validated by both exhaustive numerical simulations (using an adaptive finite element approach based on the augmented Lagrangian method), and also the previously published data. The considered porous media are constructed by randomised obstacles with various topologies; namely, square, circular and alternatively polygonal obstacles which are mimicked based on Voronoi tessellation of circular cases. Moreover, computations for the bi-dispersed obstacle cases are performed which further demonstrate the validity of the proposed universal scaling.
\end{abstract}

\section{Introduction}\label{sec:intro}

Yield-stress fluid flows through porous media are inherent to many industries including filtration, oil \& gas and mining \citep{frigaard2017bingham} and also numerous other applications such as biomedical treatments \citep{keating2003minimal}. Although in the case of Newtonian fluids many aspects of flows in porous media are well-discussed in the literature, when it comes to yield-stress fluids, our understanding of the phenomenon is limited mainly because modeling this problem is cumbersome due to the computational costs and/or the complexity of the experiments needed to carry out the analysis.

To overcome these barriers, several studies focused on pore-scale features of this problem \citep{bleyer2014breakage,shahsavari2016mobility,de2018elastoviscoplastic,bauer2019experimental,waisbord2019anomalous,chaparian2019porous}, however, it is yet unclear how to link/upscale the studies in micro-scale to macro-scale, especially due to the non-linearity of the constitutive equations which renders the bulk transport properties unpredictable from pore-scale dynamics. Nevertheless, in the intricate transport mechanism of yield-stress fluids through porous media, several mutual features can be identified regardless of the scales on which the previous studies are focused. In a number of studies \citep{talon2013determination,liu2019darcy,chaparian2021sliding,talon2022determination}, four regimes are detected in terms of flow rate ($Q$)-applied pressure gradient ($\Delta P/L$): (i) When the applied pressure gradient is less than a critical pressure gradient ($\Delta P_c/L$), there is no flow ($Q=0$); (ii) If the applied pressure gradient slightly exceeds the critical value, the flow is extremely localised in a channel and the flow rate linearly scales with the excessive pressure gradient where other parts of the fluid are quiescent; (iii) The third regime emerges when the applied pressure gradient increases, more and more channels appear (moderate values of pressure gradient) and the flow rate scales quadratically with the excessive applied pressure gradient; (iv) Finally when the applied pressure gradient is much higher than the critical value, the flow rate again scales linearly with the excessive pressure gradient.

Although these generic features/scales have been evidenced in a large number of studies, still the lack of an inclusive Darcy type expression for bulk properties is evident. The very first step for finding such a generic model is to thoroughly understand the pressure gradient threshold and more generally the {\it yield limit} which scaffolds any further progression of this aim.

In spite of the previous efforts to address the yield limit of the current problem \citep{liu2019darcy,chaparian2019porous,fraggedakis2021first}, mostly the findings are case dependent, thereby limiting their application for more complicated practical systems. As discussed, in this limit the flow is extremely heterogeneous, hence, pore-scale studies are not fully reliable since they do not contain any statistical data in ``real" porous media where a wider range of length scales are involved. Thus, in the present study, the aim is to derive a theoretical model based on yield-stress fluid flows principles and then validate the proposed model with exhaustive simulations.

To this end, we construct our porous media by randomly distributed obstacles of various shapes and lengths to avoid any biased results. Namely, three major types of obstacles are considered: circles, squares and polygons. Then, fluid flow simulations based on the adaptive augmented Lagrangian approach \citep{glowinski2011numerical,roquet2003adaptive} are performed which is shown to be a reliable tool for investigating the present problem, especially at the yield limit where non-regularised rheology is essential \citep{frigaard2005usage}. To be fit for purpose, both mono-dispersed and bi-dispersed systems are considered. We have recently delved into mono-dispersed circular obstacles by the means of pore-network approaches where a large data set has been generated \citep{fraggedakis2021first}. This data set is adopted here in conjunction with the present computational data for further validation of the proposed theory.

The outline of the present paper is as follows: the problem is described in \S \ref{sec:math} and the details of the utilized numerical method and porous media construction are highlighted. The numerical results are depicted in \S \ref{sec:examples}. The theory is developed in \S \ref{sec:scale} and the comparison with the computational results is performed. Conclusions are drawn in \S \ref{sec:conclusion}.

\section{Problem description}

\subsection{Mathematical formulation}\label{sec:math}

We consider incompressible two-dimensional Stokes flow through a set of obstacles (i.e.~$X$) in a box of size $L \times L$ (i.e.~$\Omega$) which is governed by,
\begin{equation}\label{eq:gov}
0 = -\boldsymbol{\nabla} p + \boldsymbol{\nabla} \boldsymbol{\cdot} \boldsymbol{\uptau}~~\&~~\boldsymbol{\nabla} \boldsymbol{\cdot} \boldsymbol{u}=0~~\text{in}~\Omega \setminus \bar{X},
\end{equation}
where $p$, $\boldsymbol{\uptau}$, and $\boldsymbol{u}$ represent the pressure, deviatoric stress tensor, and the velocity vector of the fluid, respectively. We use the Bingham model to describe the fluid's rheology,

\begin{equation}\label{eq:const}
\left\{
\begin{array}{ll}
\boldsymbol{\uptau} = \left( 1 + \displaystyle{\frac{B}{\Vert \dot{\boldsymbol{\upgamma}} \Vert}} \right) \dot{\boldsymbol{\upgamma}} & \mbox{iff}\quad \Vert \boldsymbol{\uptau} \Vert > B, \\[2pt]
\dot{\boldsymbol{\upgamma}} = \boldsymbol{0} & \mbox{iff}\quad \Vert \boldsymbol{\uptau} \Vert \leqslant B,
\end{array} \right.
\end{equation}
in which $\dot{\boldsymbol{\upgamma}}$ is the rate of strain tensor (i.e.~$\boldsymbol{\nabla} \boldsymbol{u} + \boldsymbol{\nabla} \boldsymbol{u}^T$) and $\Vert \cdot \Vert$ is the second invariant of the tensor. Therefore, yielding obeys the von Mises criterion.

\begin{figure}
\begin{center}
\includegraphics[width=0.5\textwidth]{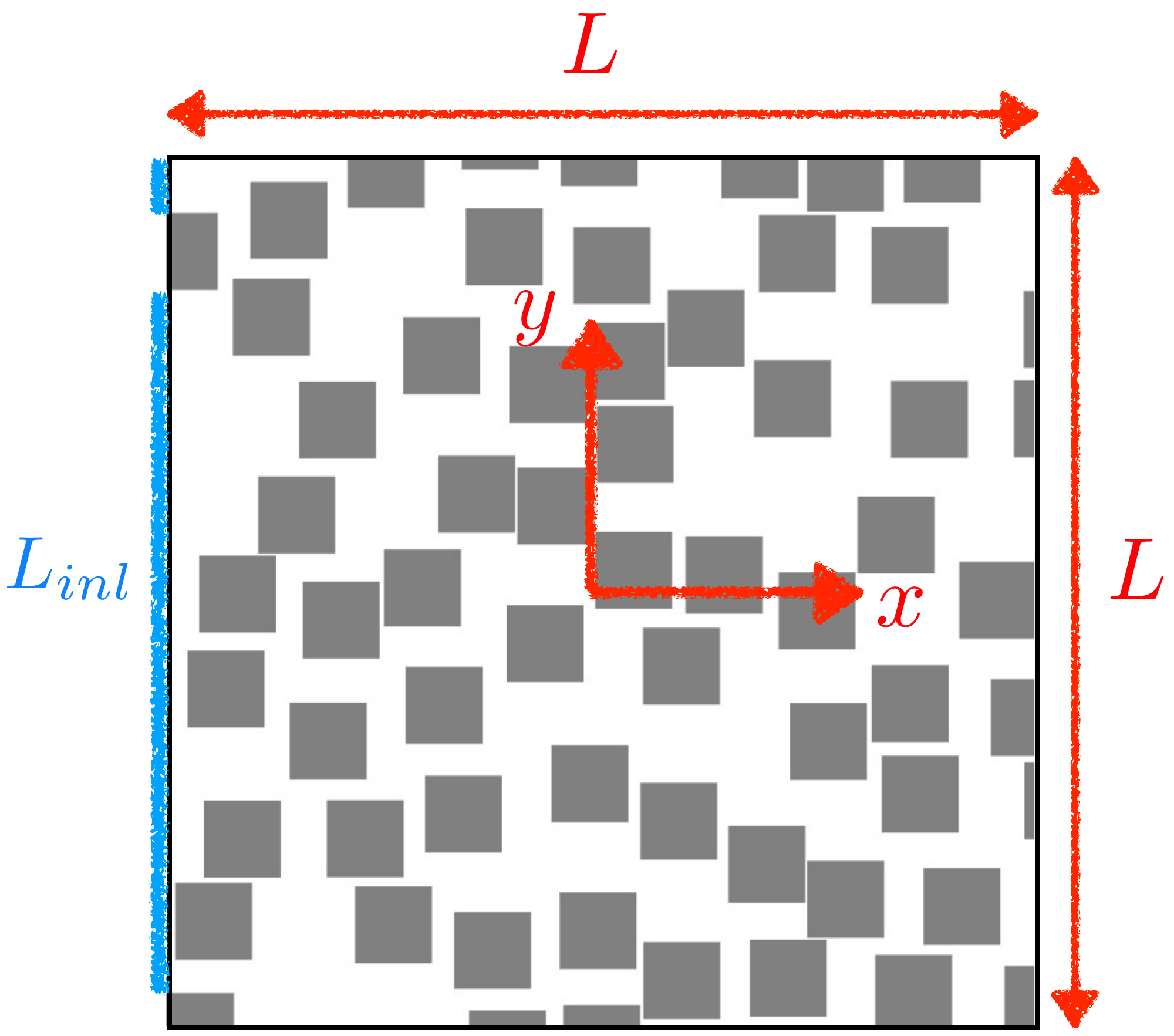}
\caption{Schematic of the coordinate system directions and the inlet length $L_{inl}$ which in this case consists of two segments depicted in blue.}
\label{fig:coordinates}
\end{center}
\end{figure}

The above equations are non-dimensional and $B=\hat{\tau}_y \hat{\ell} / \hat{\mu} \hat{V}$ is the Bingham number, where $\hat{\mu}$ is the plastic viscosity of the Bingham fluid, $\hat{V}$ is the mean inlet velocity and $\hat{\ell}$ is the characteristic length scale which will be fixed later in \S \ref{sec:PorousConstruction}. Hence, the Bingham number is the ratio of the yield stress of the fluid to the characteristic viscous stress. To derive the equations (\ref{eq:gov}) and (\ref{eq:const}), we use the following scalings:
\[
\left( x,y \right) = \frac{\left( \hat{x},\hat{y} \right)}{\hat{\ell}},~~\boldsymbol{u}=\left( u,v \right) = \frac{(\hat{u},\hat{v})}{\hat{V}}~~\&~~\left( p,\boldsymbol{\uptau} \right)=\frac{\left(\hat{p},\hat{\boldsymbol{\uptau}}\right)}{\hat{\mu} \hat{V}/\hat{\ell}},
\]
where $x$ and $y$ are the coordinates in the streamwise and spanwise directions, respectively (see figure \ref{fig:coordinates}). Please note that all the variables with {\it hat} are dimensional throughout the paper; the same symbols are used for the dimensionless parameters without $\hat{\cdot}$.

As mentioned above, $\hat{V}$ is the mean inlet velocity, hence,

\begin{equation}\label{eq:Q}
\hat{V} = \frac{\int \hat{u} ~\text{d}\hat{y}}{\hat{L}_{inl}} \Rightarrow 1 = \frac{\int (\hat{u}/\hat{V}) ~\text{d}y }{(\hat{L}_{inl}/\hat{\ell})} \Rightarrow Q = \int u ~\text{d}y = L_{inl},
\end{equation}
where $Q$ is the flow rate and $L_{inl}$ is the length of the domain's inlet, i.e.~the obstructed length by the solid obstacles is subtracted from $L$ to calculate $L_{inl}$ (see figure \ref{fig:coordinates}). Therefore, in this setting, the flow rate is always equal to $L_{inl}$, irrespective of the Bingham number. This approach in formulating the present problem is called {\it Resistance formulation} or {\it [R]}. Indeed, the \textit{yield limit} in this type of problem setup moves to $B \to \infty$. We will predominantly use this approach in our following simulations and analytical derivations. Alternatively, another formulation is possible: {\it Mobility formulation} or {\it [M]}.

In {\it [M]} approach, the applied pressure gradient is used to scale the pressure and the stress tensor (i.e.~$\frac{\Delta \hat{P}}{\hat{L}} \hat{\ell}$), while the velocity vector is scaled with $\frac{\hat{\ell}^2}{\hat{\mu}} \frac{\Delta \hat{P}}{\hat{L}}$. Hence, as a result, the non-dimensional applied pressure gradient in {\it [M]} is always equal to unity.

In {\it [M]} formulation, the independent flow parameter is,
\[
Y = \frac{\hat{\tau}_y }{\hat{\ell} (\Delta \hat{P}/\hat{L})},
\]
which is known as the \textit{yield number}. Indeed, the flow rate changes as the yield number varies: it is zero when $Y \geqslant Y_c$ and increases as the yield number drops below $Y_c$ and decreases. Indeed, the yield limit in {\it [M]} is marked by $Y_c$ which is the critical yield number; if $Y < Y_c$, the applied pressure gradient is enough to overcome the yield stress resistance and the fluid flows inside the medium.

There is a one-to-one map between {\it [R]} and {\it [M]} approaches: these two distinct formulations are linked together with $Y (\Delta P/L) = B$. This makes the interpretation of the results feasible; no matter the analysis (analytical, computational, etc.) is done in {\it [R]} or {\it [M]} settings. For more detailed explanations of these two formulations in porous media flows or more general pressure-driven flows, readers are refereed to \cite{chaparian2021sliding}.

\subsection{Porous media construction}\label{sec:PorousConstruction}

\begin{figure}
\begin{center}
\includegraphics[width=\textwidth]{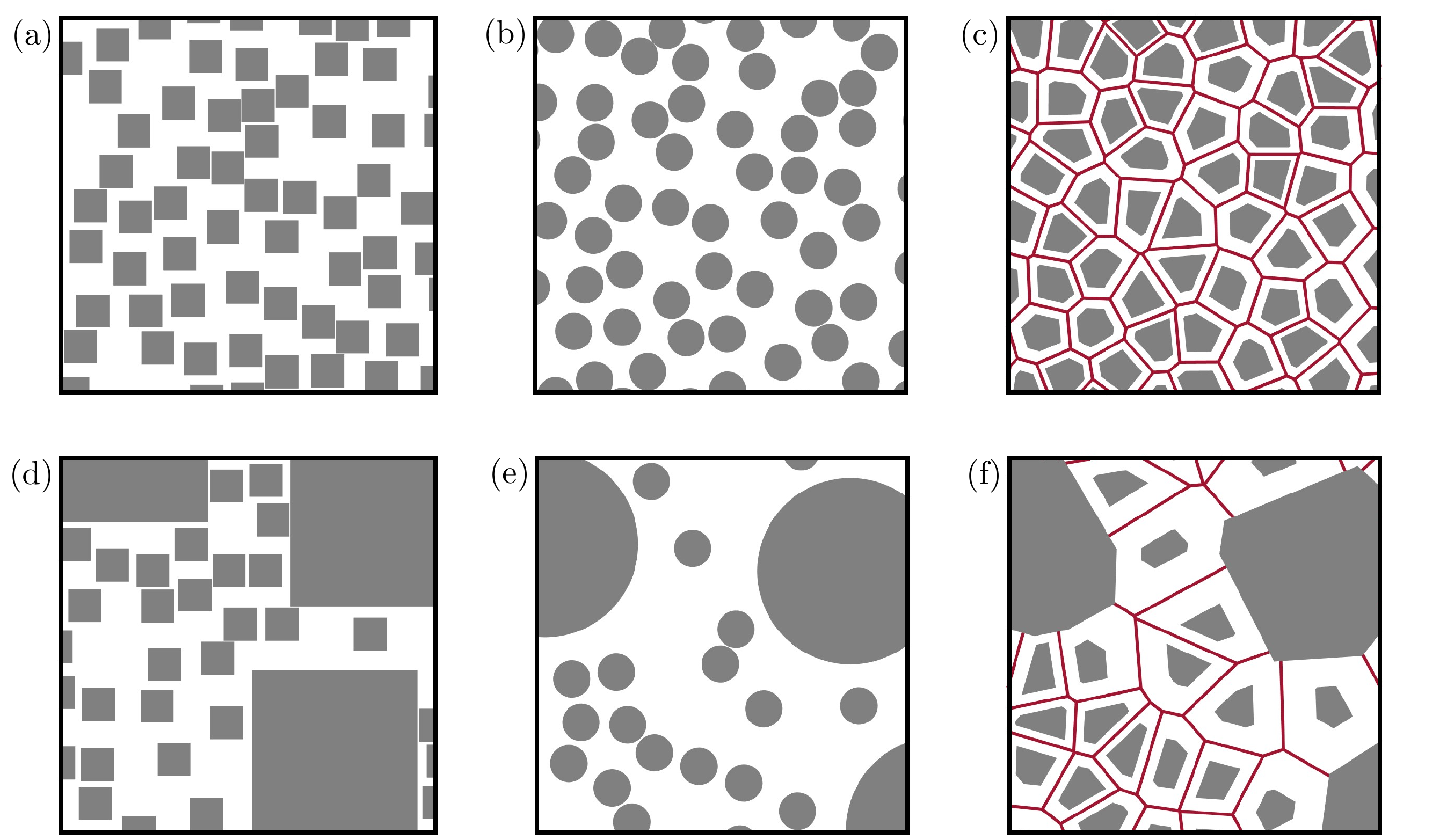}
\caption{Schematic of the porous media: top row are mono-dispersed topologies and bottom row are bi-dispersed ones. (a,d) square obstacles; (b,e) circular obstacles; (c,f) generated polygon obstacles based on Voronoi tessellation of panels (b,e).}
\label{fig:schematic}
\end{center}
\end{figure}

To construct the porous media for the fluid flow simulations, we randomly distribute non-overlapping obstacles ($X$) inside a square domain ($\Omega$) of size $L \times L = 50 \times 50$; see figure \ref{fig:schematic}. Indeed, the centre of each obstacle is chosen randomly with uniform distribution in the interval $\left[ -\epsilon, L + \epsilon \right] \times \left[ -\epsilon, L + \epsilon \right] $ and then it will be checked if the obstacle satisfies the non-overlapping condition. Here $\epsilon$ is introduced to let the obstacles cross the computational borders.

Three different obstacle topologies are used: circles, squares, and polygons. We consider mono-dispersed and bi-dispersed cases. In the mono-dispersed circular cases, the radius of the obstacles is used as the length scale, $\hat{\ell}=\hat{R}$, which deduces each individual obstacle area to be equal to $\pi$. In the mono-dispersed square cases, to be consistent, the individual area of an obstacle is again $\pi$ or indeed $\hat{\ell} = \sqrt{\hat{L}_s/\pi}$ where $\hat{L}_s$ is the length of squares' edges.

In the bi-dispersed cases, the area of the larger obstacles is $25\pi$ while the area of the smaller ones is still $\pi$. This is the only parameter which is fixed in the construction of bi-dispersed cases. To ensure generality of the results, both the positions and the number of the larger obstacles are also chosen completely randomly.

For the case of polygons (see panels (c,f) in figure \ref{fig:schematic}), firstly the domain is partitioned based on the Voronoi tessellation in which the centres of circles are adopted as the set of points in the Euclidean plane. Then each Voronoi cell (the edges of cells are depicted in red in figure \ref{fig:schematic}) is squeezed (or expanded in the bi-dispersed cases) to get the desired area of $\pi$ (or $25 \pi$ in the bi-dispersed cases). Hence, this method provides us with a variety of shapes for the polygon cases.

As mentioned, here we are interested in 2D flows, hence, the solid ``volume" fraction in the porous media is denoted by $\phi=\text{meas}(X)/\text{meas}(\Omega)$. Therefore, the porosity of the medium (i.e.~void fraction) can be represented simply by $1-\phi$.

Note that in the polygon bi-dispersed cases, the obstacles may weakly overlap because of the expansion of the cells associated with the larger obstacles. In these cases, the effective solid volume fraction is considered.

\subsection{Computational details}\label{sec:numerical}
\begin{figure}
\begin{center}
\includegraphics[width=0.75\textwidth]{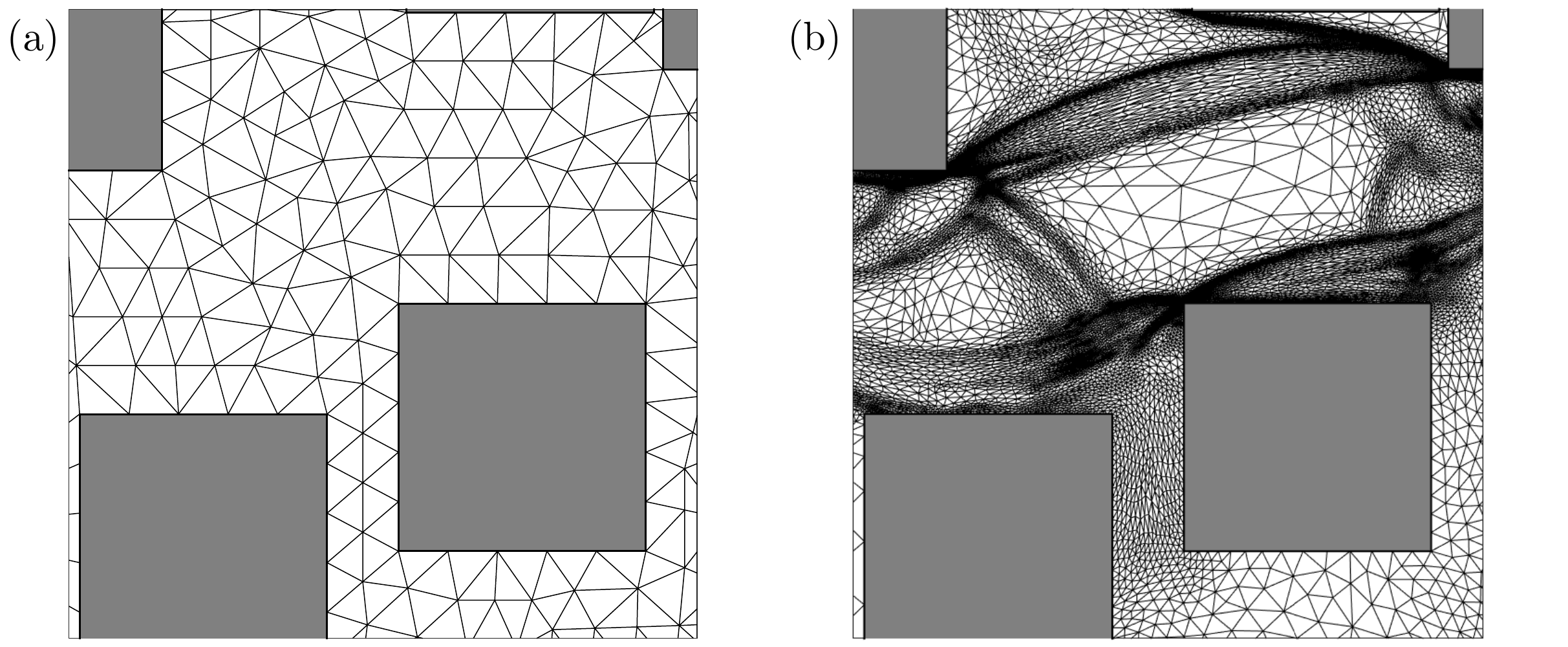}
\caption{Mesh generation for a sample case: (a) initial mesh (``uniform" coarse grid), (b) final mesh after 6 cycles of adaptation. This mesh is associated with the simulation illustrated in panel (d) of figure \ref{fig:B1000}. Note that only part of the mesh in the white window of panel (d) of figure \ref{fig:B1000} (at the pore-scale) is shown here.}
\label{fig:mesh}
\end{center}
\end{figure}

We implement augmented Lagrangian method to simulate the viscoplastic fluid flow \citep{glowinski2011numerical,roquet2003adaptive}. This method is capable of handling the non-differentiable Bingham model by relaxing the rate of the strain tensor. An open source finite element environment---FreeFEM++ \citep{MR3043640}---is used for discretisation and meshing which has been widely discussed and validated in our previous studies; for more details (choice of elements etc.) please see \cite{chaparian2017yield,iglesias2020computing,chaparian2021sliding,chaparian2022vane}. Anisotropic adaptive mesh in $\Omega/\bar{X}$ is combined with this method to get high resolution of the flow features and more accurate results. Details about the mesh adaptation can be found in \cite{roquet2003adaptive} and \cite{chaparian2019adaptive}. Basically, the initial mesh is adapted on the piecewise linear approximation of the Hessian of the square root of the dissipation function which results in finer mesh in sheared regions and coarser mesh in unyielded regions. Also the adapted mesh is stretched anisotropically to align with the yield surfaces (boundary between unyielded and yielded regions); see figure \ref{fig:mesh}.

Regarding the velocity boundary conditions: (i) no-slip is imposed at the surface of the obstacles (i.e.~$u=v=0$), (ii) natural boundary condition (or ``non-essential" finite element boundary condition) is imposed on the inlet (left face) and the outlet (right face), $\partial u/\partial x =0$, and (iii) free-slip on the top and bottom edges, $\partial u/\partial y =0$.

As discussed in \S \ref{sec:math}, in {\it [R]} setting, the flow rate must be equal to $L_{inl}$, hence, the imposed pressure gradient $\Delta P/L$ (which is a body force term in the numerical implementation) will be iterated to match the flow rate \citep{roustaei2015residuala}.

In the present study, a number of simulations are performed at different porosities to validate the scaling which will be derived in \S \ref{sec:scale}. As mentioned in \S \ref{sec:intro}, the main aim here is presenting a mathematical model for the yield limit based on physical features of the problem, and then validate it with the present simulations and the previous published data. Due to high computational cost of the full fluid flow simulations, we do not follow a statistical approach by simulating the flow in many realisations here. Rather, we use our data previously published in \cite{fraggedakis2021first} which will be discussed later in \S \ref{sec:scale}. Recent advances in computational methods of viscoplastic fluids (e.g.~known as PAL \& FISTA methods) accelerate the simulations of this type of fluids, yet the implementation of these methods is beyond the scope of this work. Interested readers are refereed to \cite{dimakopoulos2018pal} and \cite{treskatis2016accelerated,treskatis2018practical}.


\section{Illustrated examples}\label{sec:examples}
\begin{figure}
\begin{center}
\includegraphics[width=\textwidth]{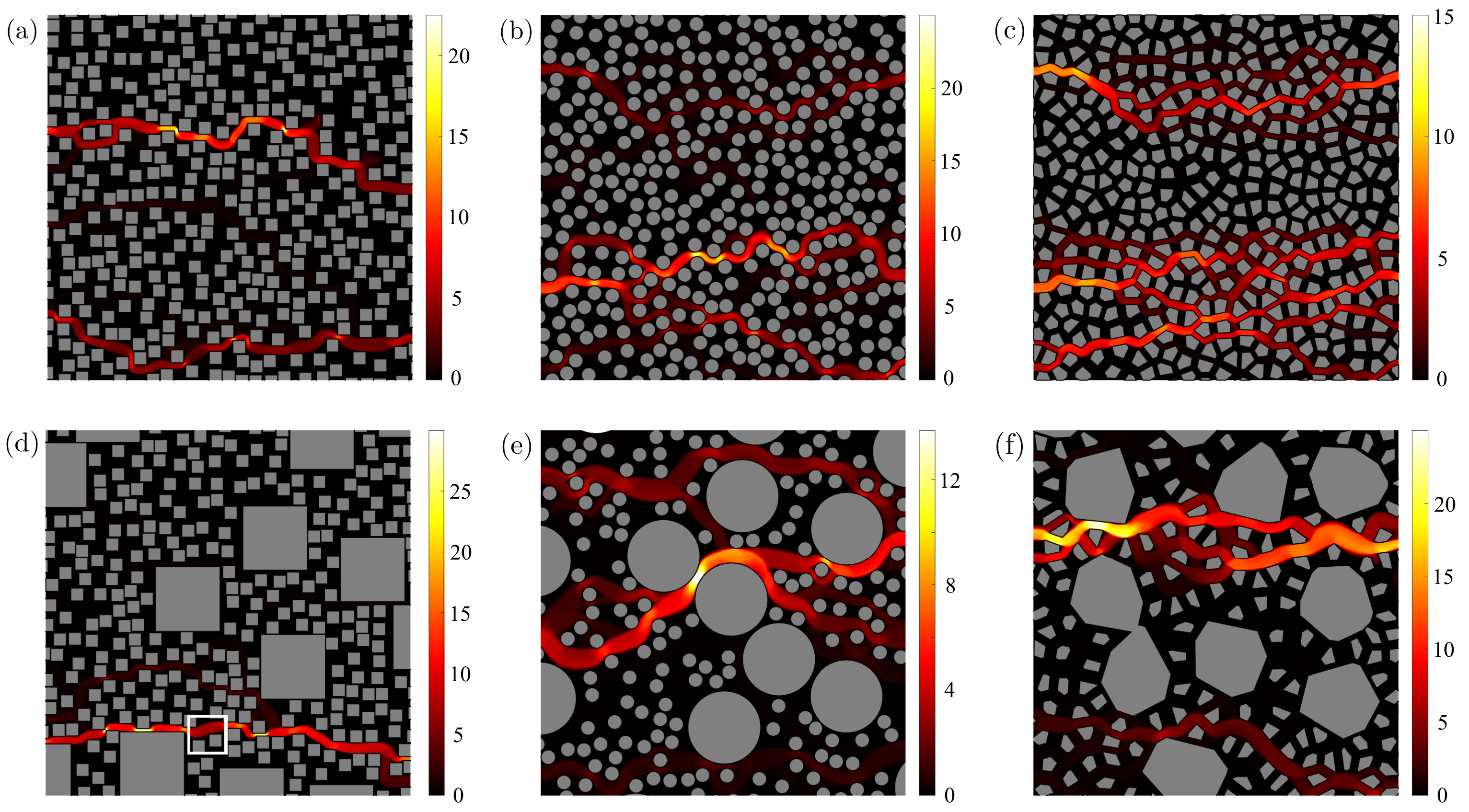}
\caption{Contour of velocity (i.e.~$\vert \boldsymbol{u} \vert$) for 6 sample simulations at $\phi=0.45~\&~B=10^3$. Top row panels are mono-dispersed cases and the bottom panels are the bi-dispersed ones. The white window in panel (d) marks where the mesh represented in figure \ref{fig:mesh} belongs to.}
\label{fig:B1000}
\end{center}
\end{figure}

Figure \ref{fig:B1000} shows the flow in the six sample geometries at $\phi=0.45~\&~B=10^3$. As discussed in \S \ref{sec:math}, the yield limit in the {\it [R]} setting goes to $B \to \infty$, so in the illustrated examples for this relatively large Bingham number, the channelisation is clear. However, clearly for different geometries, different ``large" Bingham numbers are required to get only the very first open channel. This translates to different critical yield numbers which is expected for different topologies and will be discussed in \S \ref{sec:scale} with other features of the flows.


\section{Universal scale}\label{sec:scale}

For the present problem defined in \S \ref{sec:math}, the energy equation at the steady state implies that the work done by the applied pressure gradient (i.e.~$ (\Delta \hat{P}/\hat{L}) \int_{\Omega \setminus \bar{X}} \hat{\boldsymbol{u}}  ~\text{d}\hat{A} $) balances the total dissipation (i.e.~$\int_{\Omega \setminus \bar{X}} (\hat{\boldsymbol{\uptau}} \boldsymbol{:} \hat{\dot{\boldsymbol{\upgamma}}}) ~\text{d}\hat{A} = \hat{\mu} \int_{\Omega \setminus \bar{X}} (\hat{\dot{\boldsymbol{\upgamma}}} \boldsymbol{:} \hat{\dot{\boldsymbol{\upgamma}}}) ~\text{d}\hat{A} + \hat{\tau}_y \int_{\Omega \setminus \bar{X}} \Vert \hat{\dot{\boldsymbol{\upgamma}}} \Vert ~\text{d}\hat{A}$) which in dimensionless form reads,
\[
a(\boldsymbol{u},\boldsymbol{u}) + B~ j(\boldsymbol{u}) = \int_{\Omega \setminus \bar{X}} (\dot{\boldsymbol{\upgamma}} \boldsymbol{:} \dot{\boldsymbol{\upgamma}} ) ~\text{d}A + B \int_{\Omega \setminus \bar{X}} \Vert \dot{\boldsymbol{\upgamma}} \Vert ~\text{d}A = \frac{\Delta P}{L} \int_{\Omega \setminus \bar{X}} u ~\text{d}A,
\]
where $a(\boldsymbol{u},\boldsymbol{u})$ is the viscous dissipation and $B~ j(\boldsymbol{u})$ is the plastic dissipation. At the yield limit ($B \to \infty$ in {\it [R]} or alternatively $Y \to Y_c^-$ in {\it [M]}), the viscous dissipation (which is quadratic in terms of $\dot{\boldsymbol{\upgamma}}$) is at least one order of magnitude less than the plastic dissipation \citep{frigaard2019background,chaparian2019porous}, hence, the critical yield number (or indeed the inverse of the non-dimensional critical pressure gradient) can be predicted by,
\begin{equation}\label{eq:Yc}
Y_c = \frac{\hat{\tau}_y}{\left( \displaystyle\frac{\Delta \hat{P}_c}{\hat{L}} \right) \hat{\ell}} = \lim_{B \to \infty} \frac{\displaystyle\int_{\Omega \setminus \bar{X}} u ~\text{d}A}{j(\boldsymbol{u})} 
\end{equation}

\begin{figure}
\begin{center}
\includegraphics[width=\textwidth]{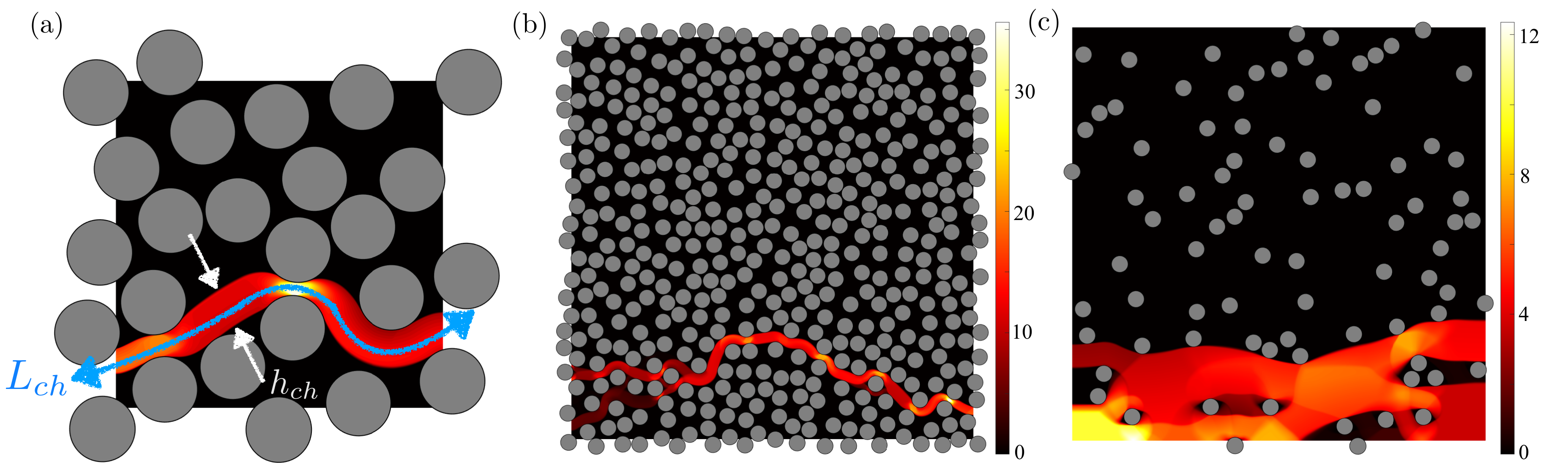}
\caption{Channelisation characteristics: (a) schematic illustration of $L_{ch}$ \& $h_{ch}$ definition, (b) velocity contour for $\phi=0.5$ \& $B=10^4$, (c) velocity contour for $\phi=0.1$ \& $B=10^4$.}
\label{fig:model}
\end{center}
\end{figure}

One can re-write the numerator as,
\begin{equation}
\int_{\Omega \setminus \bar{X}} u ~\text{d}A = L \times \int u ~\text{d}y = L \times L_{inl},
\end{equation}
since the flow rate is equal to $L_{inl}$; see expression (\ref{eq:Q}). At the yield limit, the flow in the porous media is localised to a single channel. Thus, to find the scalings for $j(\boldsymbol{u})$ and $L_{inl}$ at this limit, it is worth revisiting the two-dimensional Poiseuille flow of a yield-stress fluid. In this type of flow, the fluid moves as a core unyielded region with a constant velocity which is sandwiched between two sheared regions in which the velocity profile is parabolic. In the yield limit, these two sheared regions are viscoplastic boundary layers \citep{piau2002viscoplastic,balmforth2017viscoplastic} with thickness $\delta$. To simplify the plastic dissipation functional $j(\boldsymbol{u})$ substantially, we approximate the flow in the first open channel with the discussed Poiseuille flow. Hence, the leading order of $\Vert \dot{\boldsymbol{\upgamma}} \Vert$ can be approximated as $\approx 2 (U_{ch}/\delta) ~\delta ~L_{ch} \sim U_{ch} L_{ch}$ in the boundary layers where the index $_{ch}$ stands for the first open channel. Indeed, $U_{ch}$ and $L_{ch}$ represent the core unyielded region velocity and the length of the first channel, respectively; see figure \ref{fig:model}(a). Moreover, the continuity equation in the leading order obeys $Q= L_{inl} \approx U_{ch} h_{ch} $ which allows us to rewrite $U_{ch}$ in terms of the flow rate and the channel height. Thus,

\begin{equation}\label{eq:hoverL}
Y_c \sim \frac{L \times L_{inl}}{ U_{ch}  ~L_{ch} } = \frac{L \times L_{inl}}{(L_{inl}/ h_{ch} ) ~ L_{ch} } = \frac{ h_{ch} }{ L_{ch} /L}.
\end{equation}

In our recent study \citep{fraggedakis2021first}, we have shown that the mean height of the first open channel scales with the porosity, i.e.~$\left< h_{ch} \right> \sim 1-\phi$ and the mean relative length of the first channel scales with the volume fraction, i.e.~$\left< L_{ch} \right>/L \sim \phi$, where $\left< \cdot \right>$ stands for the mean quantity which is acquired by ensemble averaging through different simulations and also various porosities. To elaborate, in a condensed system of obstacles (i.e.~low porosities), $h_{ch}$ is smaller since the fluid path is squeezed between the obstacles or the mean void length between the obstacles becomes smaller as the solid volume fraction increases. On the other hand, the mean relative length of the first channel or {\it tortuosity} (i.e.~$L_{ch}/L$) scales with the solid volume fraction since in a denser system, the minimum path's shape is {\it zigzag} rather than straight which is more probable in a more dilute system of obstacles. These interpretations are evidenced in figure \ref{fig:model} by sample illustrations: panel (b) shows a sample simulation for $\phi=0.5$ in which the first channel is very thin and long compared to panel (c) in which $\phi=0.1$ and the channel is rather thick and straight ($L_{ch} \to L$).

Inserting the scales for the mean height and the mean relative length of the first channel to expression (\ref{eq:hoverL}), the critical yield number can be re-written as:
\begin{equation}
Y_c \sim \frac{\left< h_{ch} \right>}{\left< L_{ch} \right>/L} \sim \frac{1-\phi}{\phi} \equiv \frac{\text{Void space}}{\text{Obstructed space}},
\label{eq:us}
\end{equation}
which means that the critical yield number scales with the ratio of the void space to the solid (i.e.~obstructed) space.

\begin{figure}
\begin{center}
\includegraphics[width=0.8\textwidth]{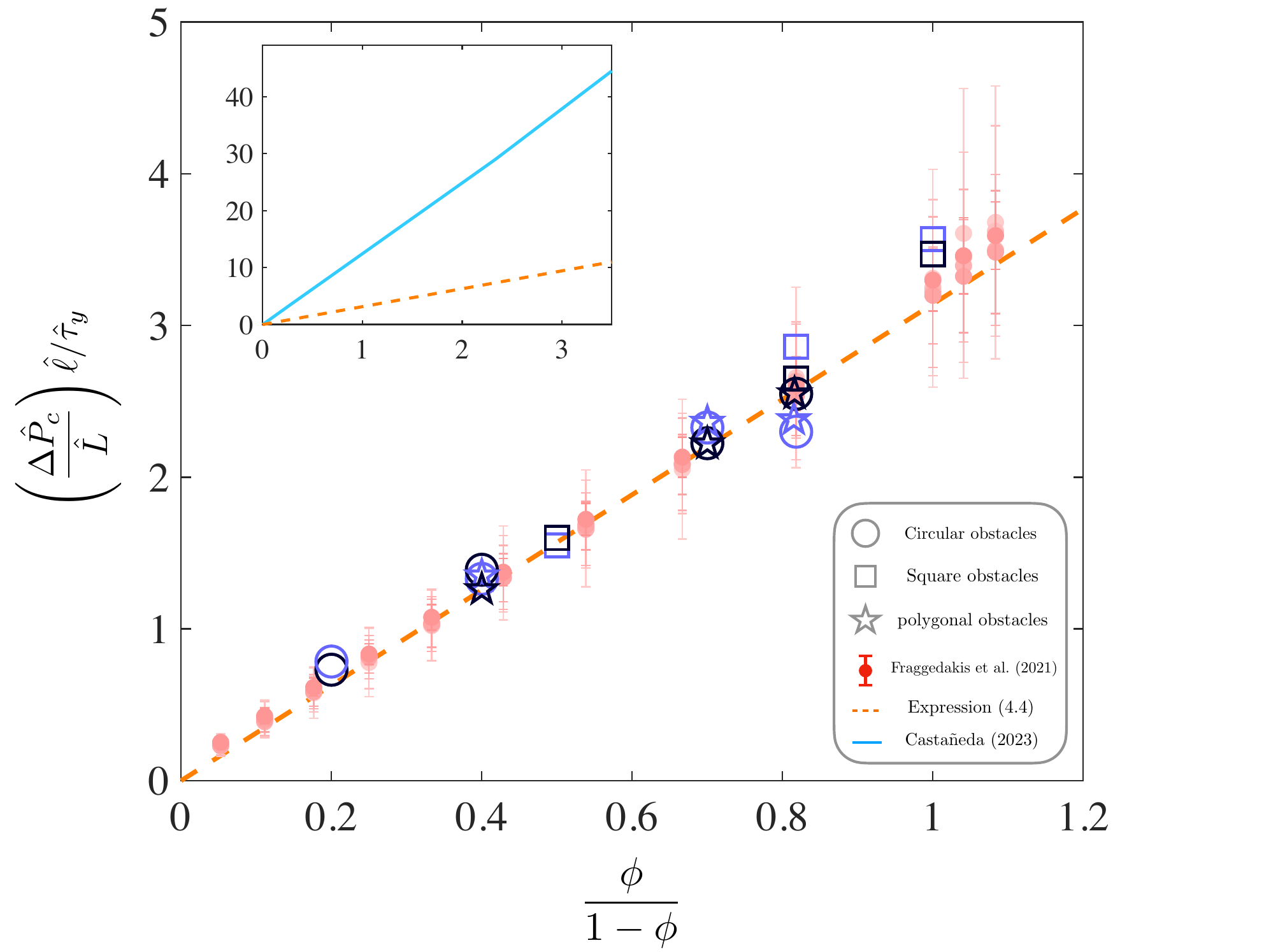}
\caption{Comparison between our theory and the computational result: non-dimensional critical pressure gradient versus $\phi/(1-\phi)$. The dashed orange line is the scale derived in (\ref{eq:Yc}). The filled circle symbols with uncertainty bars are the data borrowed from \cite{fraggedakis2021first}. Each colour intensity is dedicated to a different value of $\hat{R}/\hat{L}$ between 0.02 to 0.1 (see the reference for more details). The black and purple hollow symbols devote the mono-dispersed and bi-dispersed cases, respectively. Circles, squares, and pentagrams represent the circle, square, and polygon obstacles, respectively. Inset: comparison between the upper-bound of the critical pressure gradient (cyan line) derived by \cite{castaneda2023variational} and the proposed universal scale (dashed orange line). Please note that the axes of the inset are the same as the main figure.}
\label{fig:Yc}
\end{center}
\end{figure}

In figure \ref{fig:Yc}, we present a comparison of the theory (i.e.~expression (\ref{eq:us})) with the data associated with the simulations performed in the current study and also the previously published data: the non-dimensional critical pressure gradient (i.e.~$1/Y_c$) is plotted versus $\phi/(1-\phi)$. The dashed orange line is the scale derived above, i.e.~expression (\ref{eq:us}). The hollow symbols are the present computed data: black and purple colours are devoted to mono-dispersed and bi-dispersed cases, respectively. Circles, squares, and pentagrams represent the circle, square, and polygon obstacles, respectively. The filled circle symbols with the uncertainty bars are the data borrowed from \cite{fraggedakis2021first} where a pore-network approach is utilised to analyse a large number of realisations ($\sim 500$ for each porosity) with circular obstacles where each colour represents an specific $\hat{R}/\hat{L}$ ratio. Indeed, the filled circle symbols are the ensemble averages of all previously performed simulations and the uncertainty bars represent the range of obtained values. For more clarification of the used data, please see \cite{fraggedakis2021first}. However, as explained in \S \ref{sec:numerical}, the current data is acquired through individual simulations (i.e.~they are not ensemble averages of many simulations), hence, no uncertainty bars are associated with the new data (i.e.~hollow symbols). 

A reasonable agreement can be observed between the derived scale (with a fitted slope $\approx 3.14$ or $\pi$) and the computational data for all class of considered topologies. Moreover, the bi-dispersed cases data also fits reasonably to the proposed theory.

In a very recent study, using ``variational linear comparison" homogenisation method, \cite{castaneda2023variational} has derived an upper-bound for the critical pressure gradient where the solution of Newtonian fluids used as a test function in the dissipation-rate potential of viscoplastic fluids. This upper-bound is shown by the cyan line in the inset of figure \ref{fig:Yc} along with the proposed universal scale for comparison. Note that the upper-bound proposed by \cite{castaneda2023variational} has a linear functionality with $\phi / (1-\phi)$ which further validates the universal scale derived here, although its slope is steeper which is not surprising as it is an upper-bound.

\section{Concluding remarks}\label{sec:conclusion}

Adaptive finite element simulations based on augmented Lagrangian scheme were performed to study the fluid flows of yield-stress fluids in porous media. The specific objective was to fully understand the yield limit of this type of flows and propose a theory to address the critical applied pressure gradient which should be exceeded for flow assurance purposes. This is a vital and a very base step in proposing a generic Darcy type expression for bulk transport properties of the yield-stress fluid flows in porous media.

For this aim, and to avoid prevailing analysis, flows in various porous media constructed with a wide range of obstacle shapes are investigated. The studied geometries have been generated by randomly distributing non-overlapping obstacles of circular and square shapes. In addition, more complicated topologies (i.e.~polygon obstacles), have been generated by using the Voronoi tessellation of circular cases. The computational data includes both mono-dispersed and bi-dispersed systems.

In the yield limit, which is the main focus of the present study, the flow is restricted to a single channel connecting the inlet to outlet, while the fluid outside of it is unyielded and thus quiescent. The configuration of this very first channel has been investigated in our previous study \citep{fraggedakis2021first} and statistical geometrical properties (e.g.~height and length) are reported as a function of the solid volume fraction ($\phi$) or alternatively the porosity of the domain ($1-\phi$) which can be summarised as $\left< h_{ch} \right> \sim 1-\phi$ and $\left< L_{ch} \right> / L \sim \phi$.

A theory was proposed based on variational formulation of the energy equation. The leading order plastic dissipation has been approximated by a channel Poiseuille flow at the yield limit where the channel dimensions are borrowed from the discussed statistical results \citep{fraggedakis2021first}. Indeed, in the very first channel, the transport mechanism is predominantly postulated by the core unyielded plug in the middle of the channel and the leading order plastic dissipation occuring in the sheared boundary layer between the quiescent fluid outside of the channel and the mobilised core unyielded region. It should be noted that due to the complex shape of this limiting channel in the porous media, the mechanism is not as simple as explained above since the limiting channel is not straight and channel height varies (especially in the dense systems); see figure \ref{fig:model}. Thus, the core unyielded plug and the adjacent boundary layers are not uniform. Nevertheless, since the mean height and length of the channel is used in our model, the proposed scaling is still valid in the leading order. This has been assessed using the obtained computational data for a wide range of obstacle topologies mentioned above and also previously published data. We have shown that our theoretical approach is capable of predicting the numerical data with a reasonable agreement.

Due to the high cost of unregularised numerical simulations of yield-stress fluid flows and also handling various shapes of obstacles, the available data, especially in the yield limit, is limited. This limitation is more evident in three-dimensional flows. Although in some studies \citep{bittleston2002mud,pelipenko2004,hewitt2016heleshaw,izadi2023}, the Hele-Shaw approximation for yield-stress fluids has been developed, still the lack of a compelling study linking this pore-scale approximation to bulk transport mechanisms/features in 3D is evident. This is left for future investigations, both theoretically and computationally, which is a massive step forward for many industrial applications.


\section*{Acknowledgements}
The author thanks Babak Nasouri for fruitful discussions in the course of this study.

\section*{Declaration of interests.}The author reports no conflict of interest.

\bibliographystyle{jfm}
\bibliography{Viscoplastic.bib}

\end{document}